\documentclass[aps,prb,amsmath,amssyb,floatfix,
twocolumn,superscriptaddress,10pt]{revtex4-1}
\usepackage{graphicx}
\usepackage{subfigure}
\usepackage{hyperref}
\usepackage{braket}
\usepackage{bm}

\def\YBCO{YBa$_2$Cu$_3$O$_{6+x}$}
\def\LSCO{La$_{2-x}$Sr$_x$CuO$_4$}
\def\HBCO{HgBa$_2$CuO$_{4+\delta}$}
\def\subMF{\scalebox{0.66}{MF}}
\def\etal{{\it et al. }}
\def\ie{{\it i.e.}}
\def\bnabla{\bm{\nabla}}
\def\rhos{\rho_s}
\def\br{{\bf r}}
\def\bA{{\bf A}}
\def\bR{{\bf R}}
\def\bk{{\bf k}}
\def\bq{{\bf q}}
\def\bQ{{\bf Q}}
\def\D{\mathcal{D}}
\def\Tr{{\rm Tr}}
\def\prl{{Phys. Rev. Lett. }}
\def\prb{{Phys. Rev. B }}
\def\rmp{{Rev. Mod. Phys. }}
\def\science{{Science }}

\def\natphys{{Nat. Phys. }}
\def\natcomm{{Nat. Commun. }}

\begin{document}

\title{Long-range order and pinning of charge-density waves in
competition with superconductivity}

\author{Yosef~Caplan}
\affiliation{Racah Institute of Physics, The Hebrew University,
  Jerusalem 91904, Israel}
\author{Gideon~Wachtel}
\affiliation{Racah Institute of Physics, The Hebrew University,
  Jerusalem 91904, Israel}
\affiliation{Department of Physics, University of Toronto, Toronto,
Ontario M5S1A7, Canada}
\author{Dror~Orgad}
\affiliation{Racah Institute of Physics, The Hebrew University,
  Jerusalem 91904, Israel}

\date{\today}

\begin{abstract}

Recent experiments show that charge-density wave correlations are prevalent
in underdoped cuprate superconductors. The correlations are short-ranged at
weak magnetic fields but their intensity and spatial extent increase rapidly
at low temperatures beyond a crossover field. Here we consider the possibility
of long-range charge-density wave order in a model of a layered system where
such order competes with superconductivity. We show that in the clean limit,
low-temperature long-range order is stabilized by arbitrarily weak magnetic fields.
This apparent discrepancy with the experiments is resolved by the presence
of disorder. Like the field, disorder nucleates halos of charge-density wave,
but unlike the former it also disrupts inter-halo coherence, leading to a correlation
length that is always finite. Our results are compatible with various experimental
trends, including the onset of longer range correlations induced by inter-layer
coupling above a characteristic field scale.


\end{abstract}


\maketitle

\section{Introduction}

The pseudogap state of the cuprate high-temperature superconductors
(HTSCs) harbors various fluctuating electronic orders.\cite{intertwined}
In particular, recent nuclear magnetic resonance (NMR)
\cite{NMR-nature, NMR-nature-comm,NMR-arxiv} and x-ray scattering
\cite{Ghiringhelli,Chang,Achkar1,Blackburn13,Achkar2, Comin1,daSilva,
Le-Tacon,Hucker,Blanco-Canosa,Tabis,Croft,SilvaNeto,Comin3,Comin-symmetry,Gerber}
measurements have found evidence of charge-density wave (CDW) fluctuations across
this family of materials. The observed strength of the CDW fluctuations is
anti-correlated with superconductivity (SC) in the sense that the
intensity of the CDW scattering peak grows as the temperature is
reduced towards the superconducting transition temperature, $T_c$, and
then decreases or saturates upon entering the SC phase. In addition,
the CDW signal is enhanced when a magnetic field is used to quench SC,
while the effect of a magnetic field above $T_c$ is negligible.
Finally, optical excitation of apical oxygen vibrations promotes transient
superconducting signatures in \YBCO,\cite{Hu,Kaiser} resembling 
similar results in La$_{1.675}$Eu$_{0.2}$Sr$_{0.125}$CuO$_4$,\cite{Fausti} where 
they were conjectured to be a consequence of the melting of charge stripe order.\cite{Forst}

Motivated by these findings, Hayward \etal~\cite{Hayward1,Hayward2}
proposed a phenomenological non-linear sigma model (NLSM), which
formulates the competition between fluctuating SC and CDW order
parameters. Similar models emerge also from more microscopic
considerations.\cite{Metlitski, Efetov, Meier, Einenkel-vortex} Using
Monte-Carlo simulations, Hayward \etal~calculated the temperature
dependence of the x-ray structure factor in the absence of a magnetic
field and showed that it exhibits a maximum slightly above the
Berezinskii-Kosterlitz-Thouless temperature, $T_{BKT}$, of their
two-dimensional model. The fact that a similar peak appears in
zero-field x-ray scattering from \YBCO \cite{Ghiringhelli,Chang,Achkar1}
and \LSCO,\cite{Croft} was taken as an encouraging sign that the NLSM
is able to capture salient features of the data. The situation, however,
is more complicated and the structure factor of \HBCO\cite{Tabis} shows
no such peak. It is therefore interesting to explore the extent to which
one can reproduce and understand the various trends revealed by experiments
from the perspective of a simple model of competing orders. In particular,
we would like to ask this question with regard to the transition from short-range
correlations at low magnetic fields to longer range order at high fields,
as detected by NMR\cite{NMR-nature-comm}, ultrasound\cite{ultrasound}
and most recently x-ray scattering\cite{Gerber} measurements.


To this end we incorporate into the NLSM of Ref. \onlinecite{Hayward1}
three additional ingredients that are important for comparison with
experiments, namely, inter-layer couplings, a magnetic field and
random pinning potentials. We analyze their effects on CDW signatures
and ordering tendencies via a large-$N$ approximation,
previously used by us to study the consequences of thermally
excited vortices in the NLSM.\cite{nlsm-vortex} Averages over disorder are calculated
with the replica method, and emphasis is put on the low-temperature SC phase where
the effects of the additional factors are significant. We also present
complementary results of Monte-Carlo simulations, which we use to study
the model beyond the limits of our analytical approach.


We show that in a clean system, without a magnetic field, the
competition with SC establishes a threshold inter-layer coupling for
the stabilization of long-range CDW order. On the other hand, in the
presence of a magnetic field, any small inter-layer coupling suffices
to induce long-range order between the CDW regions which nucleate
around the cores of the Abrikosov vortices. These results are also
reflected in the low-temperature CDW structure factor of weakly
coupled layers. While it vanishes linearly with decreasing temperature
in the field-free system, it diverges when a field is present. Both
behaviors are inconsistent with the x-ray data.

In contrast, qualitative agreement with the experimental phenomenology
is obtained when the effects of a random pinning potential are
taken into account. Since favorable disorder configurations nucleate
CDW regions that survive the competition with SC down to zero temperature,
the structure factor attains a non-zero finite value in this limit.
This value grows with magnetic field, which adds vortices as CDW nucleation
centers, but true long-range phase order between the CDW regions, predicted
by mean-field theory, is avoided due to the Imry-Ma argument.\cite{Imry-Ma}
Nevertheless, we find that as the field is increased through the failed
mean-field transition, the correlation length is significantly enhanced by
the effects of inter-layer couplings. At higher temperatures the structure
factor exhibits a maximum close to $T_c$, which is washed away by both
magnetic field and stronger disorder strength, while at even higher
temperatures it becomes magnetic field-independent. The crossover field
changes little until somewhat below $T_c$, where it diverges. However,
the increase in the correlation length across it diminishes with temperature.
The reflection of these trends in experiments indicates that both order
competition and disorder are crucial elements in understanding the physics
of underdoped cuprates.

In the next section we introduce the model and lay out our findings while
moving up in complexity from the single clean layer to a disordered system
of coupled layers. The technical derivations of the results are relegated to
the appendices. In the final section we discuss our work in view of
experiments done in underdoped cuprates.

\section{Model and Results}

\subsection{The clean, single layer NLSM}

We begin with the model
considered by Hayward \etal,\cite{Hayward1} for a real 6-dimensional
order parameter, equivalent to a complex SC field $\Psi=n_1+in_2$ and
two complex CDW fields, $\Phi_x=n_3+in_4$ and $\Phi_y=n_5+in_6$. Here,
for the sake of simplicity, we disregard quartic and anisotropic CDW
terms, which appear in Ref. \onlinecite{Hayward1}, and follow our
previous strategy\cite{nlsm-vortex} of using a saddle-point
approximation for the CDW fields, which is formally justified when
their number is large.  Thus, we analyze a system described by a
complex SC field $\{\psi,\psi^*\}$, and $N-2$ real CDW fields
$\{n_\alpha\}$, where $\alpha=1\dots N-2$, whose Hamiltonian is
\begin{eqnarray}
  \label{eq:H0}
  H_0[\psi,n_\alpha] & = & \frac{1}{2}\rhos\int d^2r\,
  \Big\{|(\bnabla+2ie\bA)\psi|^2
  \nonumber \\ & & \qquad+ \sum_{\alpha=1}^{N-2}\left[\lambda(\bnabla n_\alpha)^2
  +gn_\alpha^2 \right]\Big\},
\end{eqnarray}
where $\rhos$ is the stiffness of the SC order, $\lambda\rhos$ is the
corresponding quantity for the CDW components, and $g\rhos$ is the
energy density penalty for CDW ordering. We assume that some type of
order (SC or CDW) is always locally present, in the sense of its
amplitude, but that the different order parameters compete, as
expressed by the constraint
\begin{equation}
  \label{eq:cons}
  |\psi|^2+\sum_{\alpha=1}^{N-2}n_{\alpha}^2=1.
\end{equation}
The free energy $F_0$ is given by
\begin{eqnarray}
  \label{eq:F0}
  e^{-\beta F_0} & = & \int\D\psi^*\D\psi\D n\,\delta\left(|\psi|^2
    +\sum_{\alpha=1}^{N-2}n_\alpha^2-1\right)e^{-\beta H_0}\nonumber \\
  & = & \int\D\psi^*\D\psi\D n \D\bar\sigma\,
  e^{-\beta H_0} \nonumber \\ & & \qquad\qquad
  \times e^{\frac{1}{2}\beta\rhos\int d^2r \,i\bar\sigma\left(|\psi|^2
      +\sum_\alpha n_\alpha^2-1\right)},
\end{eqnarray}
where $\beta = 1/T$ is the inverse temperature. In the limit
$N\to\infty$ we integrate over $n_\alpha$ while assuming that the
Lagrange multiplier field, $\bar\sigma$, attains its saddle-point
configuration, $\bar\sigma=i\sigma$. Since we are focusing on
signatures of the CDW deep inside the SC phase, $T\ll T_{BKT}$, we
also assume that the SC fields $\psi,\psi^*$ take their saddle-point
configurations.  Within this approximation\cite{fn-Demler}, the free
energy of the clean layer is given by
\begin{eqnarray}
  \label{eq:Trln}
  \beta F_0 & = & \frac{N-2}{2}\Tr\ln\left[\frac{1}{2}\beta\rhos\left(
      -\lambda\nabla^2+ g+\sigma\right)\right] \nonumber \\
  & & + \frac{1}{2}\beta\rhos  \int d^2r\left[|(\bnabla+2ie\bA)\psi|^2
    +\sigma\left(|\psi|^2-1\right)\right], \nonumber \\
\end{eqnarray}
where the fields $\psi,\psi^*$ and $\sigma$ are determined by the coupled
saddle-point equations
\begin{equation}
  \label{eq:SPpsi}
  \frac{\delta\,\beta F_0}{\delta\psi^*(\br)} = \frac{1}{2}\beta\rhos
  \left[-(\bnabla
  +2ie\bA)^2+\sigma\right]\psi=0,
\end{equation}
and
\begin{eqnarray}
  \label{eq:SP}
  \frac{\delta\,\beta F_0}{\delta\sigma(\br)} & = &
  \frac{N-2}{2}\Tr\left[\left(-\lambda\nabla^2 +g+\sigma\right)^{-1}
    \delta_\br\right]\nonumber \\ & & +
  \frac{1}{2}\beta\rhos\left(|\psi|^2-1\right)=0,
\end{eqnarray}
with $\delta_\br(\br',\br'')=\delta(\br'-\br)\delta(\br''-\br)$.

We first consider the case of zero magnetic field, where the SC
field assumes a uniform configuration, $\psi(\br)=\psi_0$, and
$\sigma=0$. By substituting this solution in equation (\ref{eq:SP}) we
find that
\begin{equation}
\label{eq:SP-psi0}
|\psi_0|^2=1-\frac{T}{T_{MF}},
\end{equation}
where the mean-field transition temperature is given by
\begin{eqnarray}
\label{eq:TMF}
\frac{\rhos}{T_{MF}} & = & \frac{N-2}{\lambda}{\rm Tr}\left[\left(
      -\nabla^2+g/\lambda\right)^{-1}\delta_{\br}\right] \nonumber \\
      &\simeq& \frac{N-2}{4\pi\lambda}\ln\left(\frac{32\lambda}{ga^2}\right).
\end{eqnarray}
To obtain the last expression we regularized the theory by
putting it on a square lattice of spacing $a$, and assumed $ga^2/\lambda\ll 1$.
The CDW structure factor, is defined by
\begin{equation}
  \label{eq:Sdef}
  S(\bq)=\frac{1}{L^2}\int d^2r\,d^2r'\,e^{-i\bq\cdot(\br-\br')}
  \braket{n_\alpha(\br)n_\alpha(\br')},
\end{equation}
where $\bq$ is measured from the ordering wavevector of $n_\alpha$,
$L^2$ is the layer's area, and $\braket{\cdots}$ denotes
thermal averaging. We will concentrate on the peak
value $S_{CDW}\equiv S(\bq=0)$, deferring the $\bq$ dependence to
Appendix \ref{app:abrikosov}. 
Here, the uniformity of $\psi$ and $\sigma$ readily leads to the result
\begin{equation}
  \label{eq:S1}
  S_{CDW}=\frac{T}{g\rhos},
\end{equation}
which vanishes as $T\to 0$.

The situation changes upon applying a magnetic field, $B$. The
solution of the saddle-point equations (\ref{eq:SPpsi},\ref{eq:SP}) is
expected to take the form of an Abrikosov lattice of vortices, whose
density is determined by the magnetic field. Far away from the vortex
cores, $\sigma=0$ and $\psi=\psi_0$, just as in the zero field
case. However, close to the center of each vortex, $\psi$ vanishes
linearly with the distance from the vortex center, and $\sigma$
becomes negative. As a result, there is a trapped CDW mode inside each
core, in addition to a continuum of scattering modes, which exists
also without a magnetic field. Using a tight-binding approximation
(see Appendix \ref{app:abrikosov})
for these trapped modes in equation (\ref{eq:SP}),
we can estimate their contribution to $S_{CDW}$. The result depends on the
order of limits. At a low but non-zero temperature, and $B\to 0$ (more
precisely, when $t\sim ge^{-c_1\sqrt{g\phi_0/B}}\ll gT/\rhos$), we
find that
\begin{equation}
  \label{eq:S2}
  S_{CDW} = \frac{T}{g\rhos}
  + A_1\left[1-\left(\frac{A_2}{\rhos}
  +\frac{1}{T_{MF}}\right)T\right]\frac{B}{g^2\phi_0},
\end{equation}
where $\phi_0=\pi\hbar c/e$ is the flux quantum. Here, and throughout the
paper, we denote by $A_i$,$b_i$, and $c_i$ various numerical constants.
In the other limit, of a finite magnetic
field and $T\to 0$ (when $t\gg gT/\rhos$), we obtain
\begin{equation}
  \label{eq:S3}
  S_{CDW} =  A_3\frac{T}{t\rhos}\frac{B}{g\phi_0}
  e^{c_2(t/g)\rhos(1/T-1/T_{MF})}.
\end{equation}
Therefore, the
structure factor diverges at low temperatures in the presence of a
magnetic field.

\begin{figure}[t!!!]
\centering
  \includegraphics[width=\linewidth,clip=true]{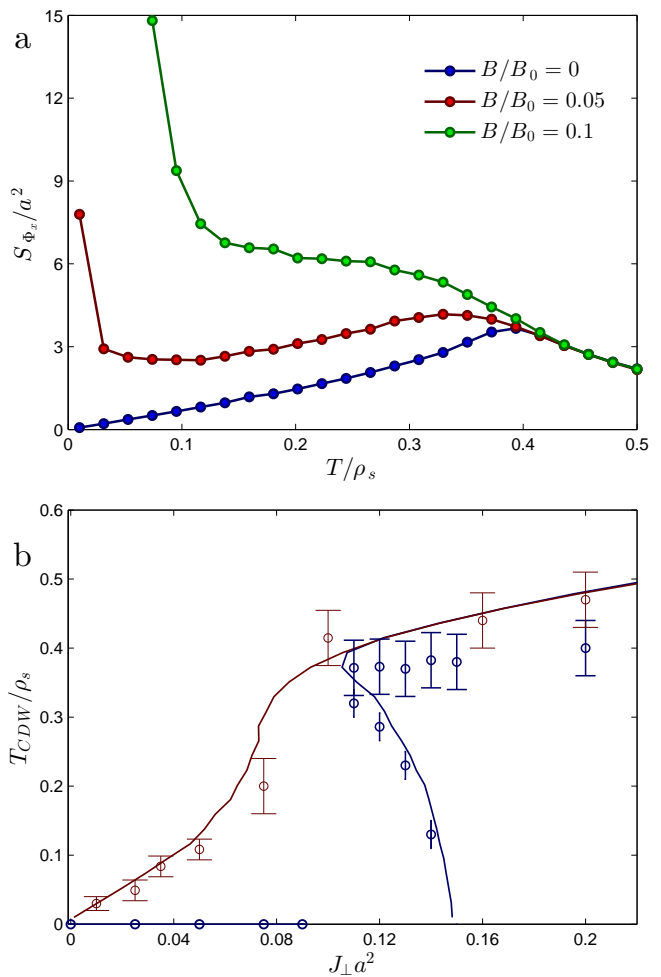}
  \caption{Onset of long-range CDW order in the clean system.
    (a) The structure factor of a clean $48\times 48$ layer with
    $\lambda=1$ and $ga^2=0.3$, as function of $T$ for three values of
    $B/B_0$, where $B_0=\phi_0/2\pi a^2$.  (b) The CDW
    ordering temperature, estimated from a $32\times 32 \times 8$
    multi-layer with the same parameters, at $B=0$ and $B=0.05B_0$. The
    lines depict the solution of the interlayer mean-field condition,
    equation (\ref{eq:mf}).}
  \label{fig:SL_clean}
\end{figure}

In order to go beyond our mean-field
results, we performed Monte-Carlo (MC) simulations of the NLSM,
incorporating the effects of a
uniform magnetic field, as appropriate for an extreme type-II
superconductor.  We used standard Metropolis updating to study
systems on a square $L\times L$ lattice, with $L$ ranging from 32 to
200 sites, and cylindrical boundary conditions. We present here MC
results for the experimentally relevant case, $N=6$, and additionally
set $\lambda=1$ and $ga^2=0.3$. For these parameters,
$T_{BKT}/\rhos=0.345$. To facilitate comparison with the results of
Ref. \onlinecite{Hayward1}, we present in Fig. \ref{fig:SL_clean}a the
structure factor $S_{\Phi_x}/a^2=2S_{CDW}/a^2$ as a function of $T$ in
a clean layer with and without a magnetic field.  For $B=0$,
$S_{\Phi_x}$ vanishes linearly as $T$ approaches zero in agreement
with equation (\ref{eq:S1}), but diverges at low $T$ for finite $B>0$, as
in equation (\ref{eq:S3}).  Our results are generally independent of the
system size, except for the $B>0,T\to 0$ limit, where the diverging
$S_{\Phi_x}$ increases with $L$.

\subsection{Clean coupled layers}

Next, we would like to ask
whether a weak interlayer CDW coupling is sufficient to stabilize
long-range CDW order. First, let us note that the diverging SC
susceptibility of each layer at $T_{BKT}$ implies that any weak
interlayer Josephson coupling of the form $-\rhos J_{SC}\int
d^2r\sum_{i}\left[\psi^*_{i}\psi_{i+1}+{\rm H.c.}\right]$, where $i$
is the layer index, induces long-range SC order. However, for weak
$J_{SC}$ the SC transition at
$T_c=T_{BKT}\left[1+b_1/\ln^2(b_2\,g/J_{SC})\right]$,
(see Appendix \ref{app:interlayer}), has only a small effect on
$|\psi|$ and thus on the amplitude and ordering tendencies of $n$.
Consequently, we concentrate on the following multi-layer Hamiltonian
\begin{equation}
\label{eq:HJperp}
H = \sum_iH_0[\psi_i,n_{\alpha,i}] -
 \rhos J_\perp \int d^2r\sum_{\alpha,i}n_{\alpha,i}n_{\alpha,i+1}.
\end{equation}
CDWs on different layers are coupled capacitively. The small amplitude of the charge
modulation associated with the CDW, and its $d$-wave nature\cite{Comin-symmetry,Fujita-symmetry}
imply a weak CDW coupling with a complicated real-space structure. We defer the study of the
consequences of such structure to a future publication and instead treat here the simplest
model interaction, as expressed in Eq. (\ref{eq:HJperp}). In the following we choose $J_\perp>0$,
although a purely repulsive interaction corresponds to $J_\perp<0$. However, the two cases are
related by the transformation $n_{\alpha,i}\rightarrow(-1)^i n_{\alpha,i}$, which reverses the
sign of $J_\perp$ but leaves $H_0$ unchanged. Consequently, our conclusions regarding
the presence of long-range CDW order hold for both attractive and repulsive interactions,
with the only difference being a change in the $c$-axis ordering wave-vector from 0 to $\pi/a$.

To estimate the effect of $J_\perp$ we use the interlayer mean-field
approximation\cite{interMF,striped} (see Appendix \ref{app:interlayer}) which in
the absence of a field yields the following condition for the putative CDW
ordering transition
\begin{equation}
\label{eq:mf}
1=2\rhos J_\perp\chi(T_{CDW}),
\end{equation}
expressed in terms of the in-plane CDW susceptibility $\chi(T)$.
For a clean system $\chi(T)=S_{CDW}(T)/T$ and
equation (\ref{eq:S1}) implies that condition (\ref{eq:mf}) can be
fulfilled only if $J_\perp\geq g/2$. When this happens uniform CDW
order is established, and the interlayer coupling term in
equation (\ref{eq:HJperp}) leads to the effective modification
$g\rightarrow g-2J_\perp\langle n_\alpha\rangle^2/\langle
n_\alpha^2\rangle$. Hence, for $J_\perp\geq g/2$ and $T\rightarrow 0$
the effective $g$ turns negative, SC disappears and the system becomes
purely CDW ordered.

In the presence of a weak magnetic field and at low temperatures, the right hand side
of the mean-field condition (\ref{eq:mf}) acquires an additional factor which scales as $B/g\phi_0$.
However, more important for establishing the qualitative difference compared to
the field-free case is the low-$T$ divergence of $S_{CDW}$, equation (\ref{eq:S3}).
This means that even for $J_\perp\rightarrow 0$, long-range order between the CDW regions
around the vortex cores does set in at $T_{CDW}=c_2 t\rhos/[g\ln(c_3 t/J_\perp)]$,
and coexists with long-range SC order.

Fig. \ref{fig:SL_clean}b depicts $T_{CDW}$ obtained from the onset
temperature of the order parameter $\sum_\alpha\langle\sum_i\int d^2r\, n_{\alpha,i}\rangle^2$
in a clean $32\times 32\times 8$ layered system, as a function of the
interlayer CDW coupling $J_\perp a^2$. As expected from the above
mean-field considerations, we find a transition for $J_\perp>g/2$ and
$B=0$, and down to the lowest accessible values of $J_\perp$ when
$B>0$.  In addition, however, for $J_\perp$ slightly below $g/2$ and
$B=0$, we observe a transition to an ordered phase which vanishes at a
lower critical temperature. This behavior can be traced to a maximum,
$\chi_{\rm max}$, in $\chi=S_{CDW}/T$, which gives two solutions to
equation (\ref{eq:mf}) in the range $1/2\rhos\chi_{\rm max}<J_\perp<g/2$.

\begin{figure}[t!!!]
  \centering
  \includegraphics[width=\linewidth,clip=true]{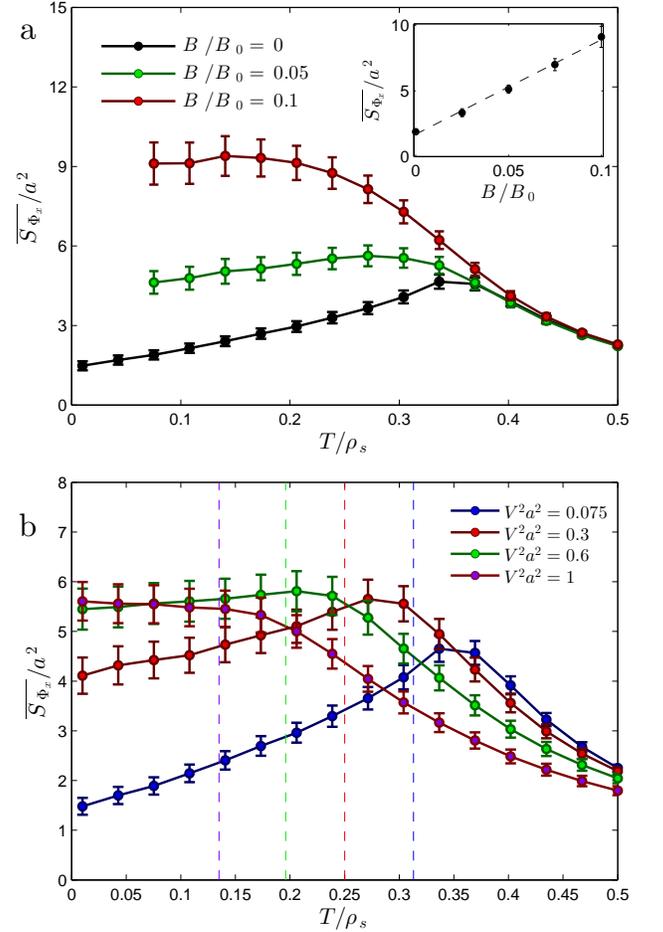}
  \caption{The structure factor of a disordered layer.
    (a) $\overline{S_{\Phi_x}}/a^2$ of a disordered $48\times 48$
    layer with $\lambda=1$, $ga^2=0.3$ and $V^2a^2=0.075$. The inset
    depicts $\overline{S_{\Phi_x}}/a^2$ as function of $B$, for
    $T=0.075\rhos$. (b) $\overline{S_{\Phi_x}}/a^2$ of the same layer at $B=0$,
    for various levels of disorder strength. The vertical lines depict $T_{BKT}$
    of each system, as deduced from the calculated superconducting phase stiffness.}
  \label{fig:disorder1}
\end{figure}

\begin{figure*}[t!!!]
  \centering
  \includegraphics[width=\linewidth,clip=true]{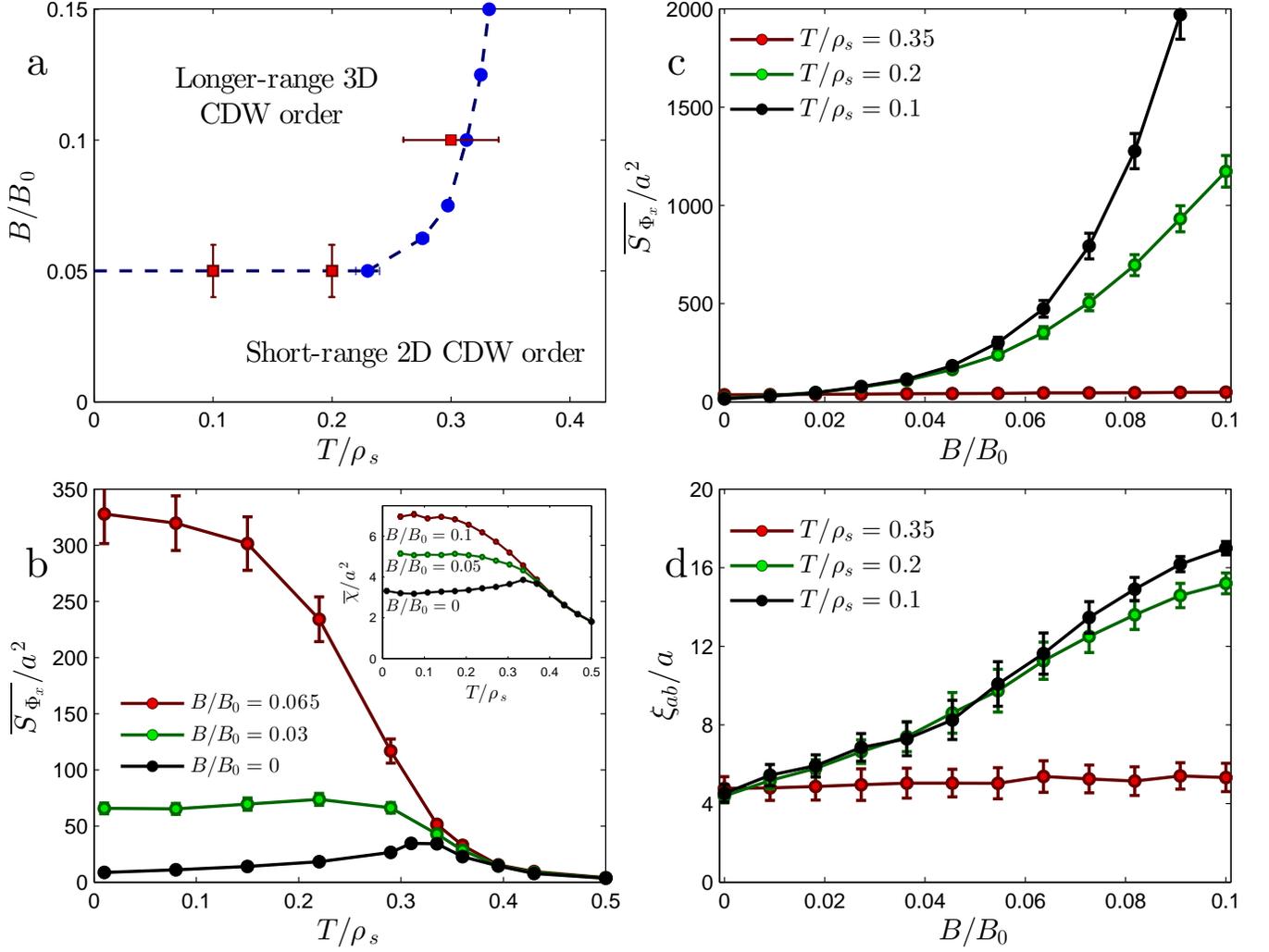}
  \caption{Crossover to longer-range CDW order in the disordered system.
  (a) The crossover line from short-range to longer-range CDW order, as deduced
  from the interlayer mean-field approximation for a
  $64\times 64\times 16$ disordered system with $\lambda=1$, $ga^2=0.3$, $V^2a^2=0.075$
  and $J_\perp a^2=0.1$. The red dots depict the onset of enhanced field and temperature
  dependence of the in-plane correlation length. (b) The structure factor, averaged over 50
  disorder realizations, as function of temperature. The inset show the CDW susceptibility
  of a single layer, used to derive the mean-field crossover line. (c) Magnetic field
  dependence of the structure factor. (d) Magnetic field dependence of the in-plane
  correlation length.}
  \label{fig:disorder2}
\end{figure*}

\subsection{Disordered single layer}

The fact that the behavior of the clean system, detailed
above, is at odds with experiments motivates us to consider the effects of a
random potential, which pins the CDW. We begin with the Hamiltonian of a
single disordered layer
\begin{equation}
  \label{eq:HV}
H_0[\psi,n_\alpha;V_\alpha] = H_0[\psi,n_\alpha] - \rhos\int d^2r\,\sum_\alpha V_\alpha n_\alpha,
\end{equation}
where $V_\alpha$ are independent random Gaussian fields satisfying
$\overline{V_\alpha}=0$ and
$\overline{V_\alpha(\br)V_\beta(\br')}=V^2\delta_{\alpha\beta}\delta(\br-\br')$,
with the overline signifying disorder averaging.

Applying the replica method to the $N\to\infty$ limit
\cite{large-n-replica,large-n-replica2} we adapt the saddle-point
equations (\ref{eq:SPpsi},\ref{eq:SP}) to the weakly disordered case, and
calculate the structure factor $\overline{S_{CDW}}$, averaged over
realizations of the pinning field. In the zero field case we find
(see Appendix \ref{app:disorder})
\begin{equation}
  \label{eq:S4}
  \overline{S_{CDW}} = \frac{T}{g\rhos} + \frac{V^2}{g^2},
\end{equation}
which decreases linearly to a finite value as the temperature is
reduced to zero. Such behavior reflects the fact that due to the
random field certain regions assume local CDW order even at
$T=0$. When the system is subject to a magnetic field,
superconductivity is suppressed inside the vortex cores, around which
CDW halos are formed. As a result, a larger fraction of the system's
area supports pinned local CDW order, and $\overline{S_{CDW}}$
increases. As long as $t\ll \max (gT/\rhos,\sqrt{gV^2})$ we find that
(see Appendix \ref{app:disorder})
\begin{eqnarray}
  \label{eq:S5}
  \!\!\!\!\!\!\!\!\!\!\!\overline{S_{CDW}}
  \!& = &\!\frac{T}{g\rhos}+\frac{V^2}{g^2} \nonumber \\
  \!& + &\!A_1\!\left[1-\left(\frac{A_2}{\rhos}
  +\frac{1}{T_{MF}}\right)T-A_4\frac{V^2}{g}\right]\!\frac{B}{g^2\phi_0}.
\end{eqnarray}
A similar functional
form characterizes the $T\to 0$ spatially averaged Edwards-Anderson
order parameter
\begin{equation}
  \label{eq:qEA}
  q_{EA} =\frac{1}{L^2}\int d^2r \overline{\braket{n_\alpha(\br)}^2}
  = \frac{V^2}{4\pi\lambda g} + A_5\frac{B}{g\phi_0}.
\end{equation}
Random-field models in the $N\to\infty$ limit
do not exhibit a glass transition\cite{large-n-replica}, and
$q_{EA}>0$ for all $T$.

In Fig. \ref{fig:disorder1}a we present results for $\overline{S_{\Phi_x}}$, averaged
over 60 disorder realizations, for a layer with $V^2a^2=0.075$. The inset
depicts its low-temperature $B$ dependence. In
accordance with our analytical result, equation (\ref{eq:S5}),
$\overline{S_{\Phi_x}}$ assumes a finite value for $T,B\to 0$, and
grows linearly with both $T$ and $B$. The error bars in our MC
results, which grow with increasing $B$ and decreasing $T$, reflect
the low convergence rates and sensitivity to initial conditions which
arise in this limit. However, our MC simulations clearly show that
$\overline{S_{\Phi_x}}$ does not diverge when $B>0$, even for
temperatures below the range presented in the figure.
In Fig. \ref{fig:disorder1}b we show $\overline{S_{\Phi_x}}/a^2$ of the
same system at $B=0$ for various disorder strengths.
The figure also depicts the
BKT transition temperature for each system, as deduced from the calculated renormalized
superconducting phase stiffness $\rhos(T)$ and the BKT criterion $\rhos(T_{BKT})=2T_{BKT}/\pi$.
Our results clearly show that the peak in the structure factor, which occurs slightly
above $T_{BKT}$, disappears with increasing disorder strength.
At the same time the zero-temperature value of $\overline{S_{\Phi_x}}/a^2$ increases and
approaches a limiting value as global superconducting order is suppressed by the disorder.

\begin{figure}[t!!!]
\centering
  \includegraphics[width=\linewidth,clip=true]{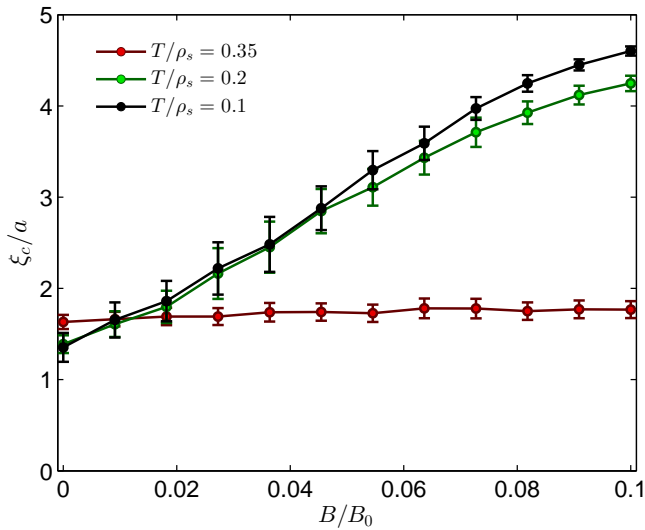}
\caption{The inter-plane correlation length.
  The magnetic field dependence of $\xi_c$ for a $64\times 64\times 16$
  disordered system with $\lambda=1$, $ga^2=0.3$, $V^2a^2=0.075$
  and $J_\perp a^2=0.1$, averaged over 50 disorder realizations.}
\label{supfig:xi_z}
\end{figure}

\begin{figure}[t!!!]
\centering
  \includegraphics[width=\linewidth,clip=true]{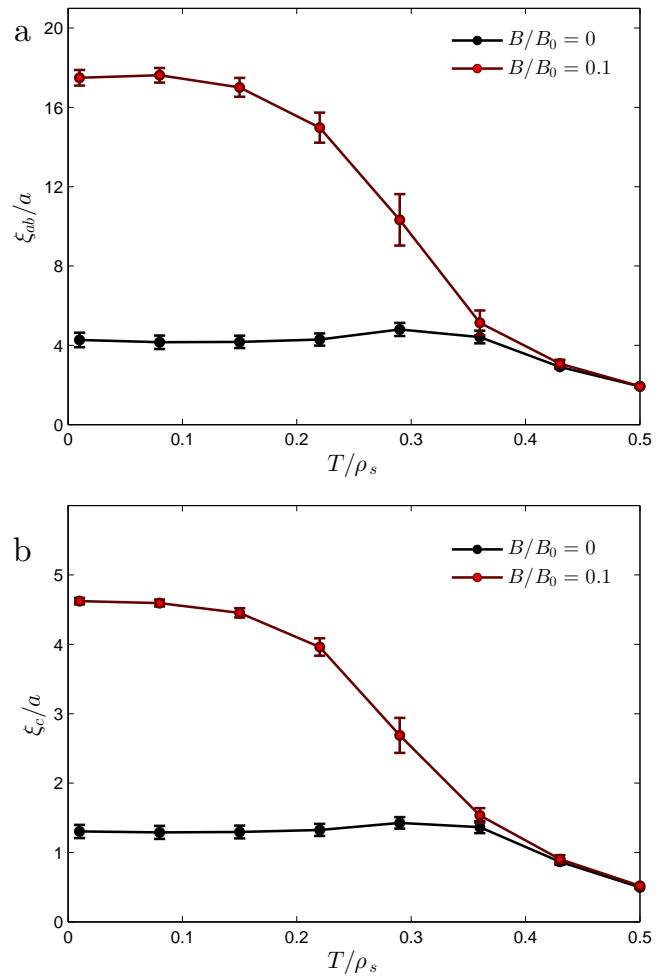}
\caption{Temperature dependence of the correlation lengths.
  $\xi_{ab}$ and $\xi_c$ for a $64\times 64\times 16$
  disordered system with $\lambda=1$, $ga^2=0.3$, $V^2a^2=0.075$
  and $J_\perp a^2=0.1$, averaged over 50 disorder realizations.}
\label{supfig:T-xi}
\end{figure}

\subsection{Coupled disordered layers}

Finally, consider the disordered version of the coupled-layer Hamiltonian (\ref{eq:HJperp}),
where each plane is described by $H_0[\psi,n_\alpha;V]$, equation (\ref{eq:HV}). While
interlayer mean-field approximation predicts a CDW ordering transition
at $T_{CDW}$, given by condition (\ref{eq:mf}), the Imry-Ma argument\cite{Imry-Ma}
precludes long-range CDW order in a disordered system below four dimensions.
Nevertheless, for weak disorder we expect the failed thermodynamic transition to leave its
mark in the form of a crossover, which signifies the onset of enhanced CDW correlations
within and between the planes.

To test this prediction, we start by evaluating the mean-field transition temperature from
condition (\ref{eq:mf}). We do so by calculating the disordered-averaged
CDW susceptibility, $\overline{\chi}$, for a layer with $\lambda=1$, $ga^2=0.3$ and $Va^2=0.075$,
which exhibits an x-ray structure factor with a similar temperature dependence to the one measured
at low fields in \YBCO,\cite{Chang} (see Fig. \ref{fig:disorder1}a). The results, presented in the
inset of Fig. \ref{fig:disorder2}b, show that $\overline{\chi}$ is approximately constant at low $T$
and grows linearly with $B$ from $1/\rhos g$, in accord with our large-$N$ analysis
(see Appendix \ref{app:disorder}). When combined with equation (\ref{eq:mf}) this implies, for
$J_\perp<\rhos g/2$, a mean-field transition that occurs at a critical field which is constant
over a wide temperature range and then increases rapidly. Such behavior, depicted in
Fig. \ref{fig:disorder2}a for a system with $J_\perp a^2=0.1$, resembles that of the transition
line into a long-range CDW phase, as deduced from ultrasound measurements.\cite{ultrasound}

Next, let us inquire what features of the mean-field transition survive the effects of fluctuations.
Fig. \ref{fig:disorder2}b depicts the temperature dependence of the structure factor (now defined by
the average of $\langle n_\alpha(\br)n_\alpha(\br ')\rangle$ over the three-dimensional system).
While the qualitative features follow the ones displayed by the two-dimensional layer,
the coupled layers exhibit, at low $T$, a rapidly increasing $\overline{S_{\Phi_x}}$ beyond a
characteristic field scale, (see Fig. \ref{fig:disorder2}c). Another relevant signature is displayed
by the in-plane correlation length, $\xi_{ab}$, defined by the inverse half-width at half maximum
of $\overline{S_{\Phi_x}}(\bq)$. Its field dependence, shown in Fig. \ref{fig:disorder2}d, exhibits
an inflection point across the same magnetic field scale. Recent x-ray measurements\cite{Gerber} have found a similar
behavior in \YBCO at a range of magnetic fields that is comparable to ours if one identifies the
short-distance scale, $a$, of the NLSM with 2-3 Cu-Cu spacings. Using also the temperature dependence
of $\xi_{ab}$ and the corresponding data for the inter-plane correlation length, $\xi_c$,
(see Figs. \ref{supfig:xi_z} and \ref{supfig:T-xi}) we can map out the crossover field in the $B-T$ plane,
and find that it follows the mean-field transition line. Hence, the following picture emerges: At weak magnetic
fields the CDW correlations are short-ranged and largely confined to the planes. However, as the field increases
at low temperatures, it combines with the inter-plane coupling to induce a crossover to more extended
three-dimensional correlations. This crossover becomes sharper with diminishing disorder.

\section{Discussion}

We have shown that competition between CDW and SC orders in the presence of
disorder, can account for many of the trends observed in x-ray scattering experiments.
Specifically, nucleation of CDW at regions of strong attractive disorder makes
$S_{CDW}$ attain a finite value at $T=0$, even for $B=0$.\cite{Ghiringhelli,Tabis,Croft}
This value increases linearly with low $B$,\cite{Chang} due to pinned CDW around vortex
cores, as seen in scanning-tunneling experiments\cite{Hoffman,Hamidian,Machida}.
At higher temperatures and for weak disorder both our simulations and experiments on
\YBCO\cite{Chang} and La$_{2-x}$Sr$_x$CuO$_4$\cite{Croft} exhibit a maximum in $S_{CDW}$
close to $T_c$, which disappears with increasing $B$, while there is practically no
$B$ dependence beyond this temperature.\cite{Chang} We find that the peak is also washed
away with increasing disorder, a fact which may explain its absence in
HgBa$_2$CuO$_{4+\delta}$.\cite{Tabis}. We note, however, that
the predicted linear low-$T$ dependence for $B=0$, (see
Fig. \ref{fig:disorder2}), is reflected in some\cite{Croft}, but not all
x-ray data.\cite{Chang} Another discrepancy exists with Ref. \onlinecite{Achkar2},
where $S_{CDW}$ was found to decrease upon increasing the amount of
oxygen disorder.

Our results also demonstrate a crossover to a regime with longer-range CDW correlations
at high magnetic fields, in accord with x-ray measurements.\cite{Gerber} Moreover, the
temperature dependence of the crossover field follows closely the one observed in
NMR\cite{NMR-nature-comm} and ultrasound\cite{ultrasound} measurements.
A commonly advocated scenario \cite{Millis-Norman,Yao,QO-Sebastian-prl,
QO-Sebastian-rep, Allais-connecting} for explaining recent quantum
oscillations experiments,\cite{QO-Doiron-Leyraud,QO-LeBoeuf,
QO-Barisic, QO-Doiron-Leyraud-small} invokes long-range CDW order
as a cause for Fermi-surface reconstruction. It is tempting to associate the
above mentioned regime of enhanced CDW correlations with this scenario.
However, the spatial extent of CDW ordering needed to explain the quantum
oscillations experiments remains to be resolved.


\acknowledgments
This research was supported by the Israel Science Foundation (Grant
No. 585/13).

\appendix
\section{CDW spectrum in the Abrikosov vortex lattice state}
\label{app:abrikosov}

Consider the NLSM for a clean layer with $B>0$ and $T<T_{BKT}$,
where an Abrikosov lattice of vortices is expected to develop.
In terms of the orthonormal eigenfunctions, $\phi_s(\br)$, and
eignevalues, $\varepsilon_s$, of the operator
\begin{equation}
  \label{eq:L}
  \hat{L}=-\lambda\bnabla^2+g+\sigma(\br),
\end{equation}
the saddle-point equations take the form
\begin{equation}
  \label{eq:SPpsisup}
  \left[-(\bnabla+2ie\bA(\br))^2+\sigma(\br)\right]\psi(\br)=0,
\end{equation}
and
\begin{equation}
  \label{eq:SPsup}
  \sum_s\frac{|\phi_s(\br)|^2}{\varepsilon_s}
  =\frac{\beta\rhos}{N-2}\left[1-|\psi(\br)|^2\right].
\end{equation}

We have previously derived an effective Ginzburg-Landau theory for the
NLSM \cite{nlsm-vortex}, and showed that the vortex core radius scales
at low temperatures as $r_0\sim g^{-1/2}$. From
Eq. (\ref{eq:SPpsisup}) it then follows that inside the core
$\sigma\sim-1/r_0^2\sim - g$, while $\sigma=0$ for $r\gg
r_0$, where $|\psi|=|\psi_0|$. Consequently, one expects that in the
presence of a vortex, the spectrum of $\hat{L}$ consists of a
continuum of scattering states with $\varepsilon_s\ge g$, and a
discrete set of bound states with $\varepsilon_s< g$. Our numerical
solution of the saddle-point equations confirms these expectations and
shows that for $\lambda=1$ there is a single bound state,
$\varphi_0(\br)$, with eigenvalue $\varepsilon_0\ll g$, which decays
at large distances as $\varphi_0(\br)\sim\exp(-r/r_0)$.

In the presence of a dilute Abrikosov lattice of vortices, \ie,~one
for which the inter-vortex distance, $R$, obeys $R\gg r_0$, the small
overlap between bound states in neighboring cores leads to the
formation of a tight-binding band $\phi_{0,\bk}(\br)$.
For a square vortex lattice its dispersion takes the form
\begin{equation}
  \label{eq:eps0k}
  \varepsilon_0(\bk)=\tilde\varepsilon_0-2t\left[\cos(k_x R)+\cos(k_y R)\right],
\end{equation}
with $|k_{x,y}|<\pi/R$, $\tilde\varepsilon_0=\varepsilon_0-\Delta\varepsilon_0$, and
\begin{eqnarray}
  \label{tight-binding}
  \Delta\varepsilon_0&=&-\int d^2r \Delta\sigma_{\bR=0}(\br)\varphi_0^2(\br), \\
  t&=&-\int d^2r \varphi_0(\br)\Delta\sigma_{\bR=R\hat{x}}(\br)\varphi_0(\br-R\hat{x})
  \nonumber\\
  &&-\Delta\varepsilon_0\int d^2r \varphi_0(\br)\varphi_0(\br-R\hat{x}).
\end{eqnarray}
Here,
\begin{equation}
  \label{dltasigma}
  \Delta\sigma_{\bR}(\br)=\sum_{\bR'\neq\bR}\sigma_V(\br-\bR'),
\end{equation}
where $\sigma_V(\br)$ is the configuration assumed by $\sigma$ in the
presence of a single vortex.  Consequently, using $R=\sqrt{\phi_0/B}$,
both $\Delta\varepsilon_0$ and $t$ scale as
$g\exp(-c_1\sqrt{g\phi_0/B})$, where $c_1$ is a constant that depends
on $\lambda$.

Under the specified conditions the scattering states still form a
continuum with $\varepsilon_s\ge g$. Since $\phi_{0,\bk}(\br)$ vanish
rapidly between vortices it follows from Eq. (\ref{eq:SPsup}) that
\begin{equation}
  \label{eq:scatstates}
  \sum_{s\in{\rm scattering}}\frac{|\phi_s(\br)|^2}{\varepsilon_s}
  =\frac{\beta\rhos}{N-2}\left[1-|\psi_0|^2-|\delta\psi(\br)|^2\right],
\end{equation}
where $\delta\psi(\br)$ is appreciable only within the cores.
Therefore,
\begin{equation}
  \label{eq:spbound}
  \sum_{\bk}\frac{|\phi_{0,\bk}(\br)|^2}{\varepsilon_0(\bk)}=
  \frac{\beta\rhos}{N-2}\left[|\psi_0|^2-|\psi(\br)|^2+|\delta\psi(\br)|^2\right],
\end{equation}
whose integral over $\br$ gives
\begin{equation}
  \label{eq:intsp}
  \int_{BZ} d^2k \frac{1}{\varepsilon_0(\bk)}={\cal C}\beta\rhos|\psi_0|^2
  \left(\frac{r_0}{R}\right)^2,
\end{equation}
with a constant ${\cal C}$. Evaluating the integral and using
$|\psi_0|^2=1-T/T_{MF}$, gives
\begin{equation}
\label{eq:intk1}
  \frac{1}{\tilde\varepsilon_0-4t}\simeq\left\{
  \begin{array}{cc}
  \!\!\frac{{\cal C}}{4\pi^2}\rhos\left(\frac{1}{T}-\frac{1}{T_{MF}}\right)r_0^2
  & :t\ll T/\rhos r_0^2 \\
  \!\!\frac{1}{32t}\exp\left[\frac{{\cal C}}{\pi}t\rhos\left(\frac{1}{T}
  -\frac{1}{T_{MF}}\right)r_0^2\right]
  & :t\gg T/\rhos r_0^2
  \end{array}
  \right.
\end{equation}

The effective action for the CDW fields, $n_\alpha$, is of the form
$(\beta\rhos/2)\int d^2r\sum_\alpha n_\alpha \hat{L} n_\alpha$, with
the result that
\begin{equation}
  \label{eq:Sq1}
  S(\bq)=\frac{1}{\beta\rhos}\frac{1}{L^2}\sum_s\frac{1}{\varepsilon_s}
  \left|\int d^2r e^{-i\bq\cdot\br}\phi_s(\br)\right|^2.
\end{equation}
Bloch's theorem implies that
\begin{equation}
  \label{eq:Bloch}
  \phi_s(\br)=e^{i\bk\cdot\br}u_{n,\bk}(\br),
\end{equation}
with $\bk$ lying within the magnetic Brillouin zone,
and $u_{n,\bk}(\br+\bR)=u_{n,\bk}(\br)$ for any position $\bR$
of the $N_V$ vortices in the lattice. Therefore,
\begin{equation}
  \label{eq:Sq2}
  S(\bq)=\frac{1}{\beta\rhos}\frac{N_V^2}{L^2}\sum_{n}\frac{1}{\varepsilon_n(\bq)}
  \left|\int_{u.c.} d^2r e^{-i\bQ\cdot\br}u_{n,\bq'}(\br)\right|^2,
\end{equation}
where the integration is over a unit cell of the Abrikosov lattice,
and where we decomposed $\bq=\bq'+\bQ$ into its projection $\bq'$ to
the first Brillouin zone and a reciprocal lattice vector $\bQ$.

Due to the spatial integration, the scattering states contribution to
$S_{CDW}=S(\bq=0)$ is dominated by the lowest lying extended state with
$\varepsilon_s\simeq g$, which is the descendent of the $\bk=0$
state of the system with $B=0$. Therefore, its integral satisfies
$(1/L^2)\left|\int d^2r \phi_s(\br)\right|^2=1-{\cal O}(r_0^2/R^2)$,
where the correction is due to its deviations from uniformity in the vicinity
of the cores. States with $\varepsilon_s>g$ provide further contributions of
order ${\cal O}(r_0^2/R^2)$. For the localized band we have
\begin{equation}
  \label{eq:u0k}
  u_{0,\bk}(\br)=\frac{1}{\sqrt{N_V}}\sum_{\bR}e^{i\bk\cdot(\bR-\br)}
  \varphi_0(\br-\bR),
\end{equation}
implying that for $R\gg r_0$
\begin{equation}
  \label{eq:intu0}
  \int_{u.c.} d^2 r\, u_{0,\bk=0}(\br)\simeq\frac{1}{\sqrt{N_V}}\int_{u.c.} d^2 r\,
  \varphi_{0}(\br)\sim \frac{r_0}{\sqrt{N_V}},
\end{equation}
since $\varphi_0(\br)$ is normalized and appreciable within $r\lesssim r_0$.
Consequently, the contribution of the band of core states to $S_{CDW}$ is of order
$(r_0/R)^2T/\rhos\varepsilon_0(\bk=0)$. Using Eq. (\ref{eq:intk1}) and
combing the two contributions, we finally arrive at Eqs. (\ref{eq:S2}) and (\ref{eq:S3}).

When applying the above considerations to a triangular Abrikosov
lattice, one needs to take into account that for this geometry
$\varepsilon_0(\bk)=-4t\left[\cos^2(k_xR/2)+\cos(k_xR/2)\cos(\sqrt{3}k_yR/2)-1/2\right]$
with $BR^2=2\phi_0/\sqrt{3}$.
However, these changes do not affect the functional dependence of
$S_{CDW}$, but only the various numerical constants, which appear in
the solution.

\section{The interlayer mean-field approximation}
\label{app:interlayer}

Here, we trade the coupled-layer problem with an effective single-layer Hamiltonian.
For the case of disordered CDW-coupled layers, the latter takes the form
\begin{eqnarray}
  \label{eq:HMFmethods}
  \nonumber
 \!\!\!\!\!\!\!\!\!\!\! H_{\subMF,i}&\!=\!&H_0[\psi_i,n_{\alpha,i};V_{\alpha,i}] \\
  &\!-\!&\rhos J_\perp \!\int d^2 r\sum_\alpha \sum_{j=i\pm 1} n_{\alpha,i}(\br)
  \langle n_{\alpha,j}(\br)\rangle_{\subMF,j}, \,
\end{eqnarray}
where $\langle\cdots\rangle_{\subMF,i}$ denotes averaging with respect to $H_{\subMF,i}$.

We are interested in the vicinity of the putative CDW ordering temperature,
$T_{CDW}$, where we would like to treat the $J_\perp$ term perturbatively. For the clean system,
this is justified by the smallness of $\langle n_{\alpha,j}(\br)\rangle_{\subMF,j}$ close to $T_{CDW}$.
In the disordered case $\langle n_{\alpha,j}(\br)\rangle_{\subMF,j}$ is random and may be large
in regions where the pinning potential is strong enough to overcome the
effects of thermal fluctuations. Hence, when disorder is present we assume that $J_\perp a^2$ is small
and obtain to leading order
\begin{eqnarray}
  \label{eq:MFcond0}
  \nonumber
 &&\hspace{-1cm}\langle n_{\alpha,i}(\br)\rangle_{\subMF,i}= \langle n_{\alpha,i}(\br)\rangle_0 \\
  &&\hspace{-0.4cm}+\rhos J_\perp \int d^2 r'\sum_\beta \sum_{j=i\pm 1} \chi_{\alpha\beta,i}(\br,\br')
  \langle n_{\beta,j}(\br')\rangle_{\subMF,j}, \,
\end{eqnarray}
where
\begin{eqnarray}
\label{eq:chidef}
\nonumber
&&\hspace{-1cm}\chi_{\alpha\beta,i}(\br,\br') \\
&&\hspace{-0.4cm}=\frac{1}{T}\left[
    \braket{n_{\alpha,i}(\br)n_{\beta,i}(\br')}_0-\braket{n_{\alpha,i}(\br)}_0\braket{n_{\beta,i}(\br')}_0\right],
\end{eqnarray}
is the in-plane CDW response function, and $\langle\cdots\rangle_0$ signifies averaging with respect to $H_0$.

Next, we average equation (\ref{eq:MFcond0}) over the disorder realizations. Since the pinning potentials on different
layers are assumed independent, any correlations between $\overline{\chi_{\alpha\beta,i}}$ and
$\overline{\langle n_{\beta,i\pm 1}\rangle_{\subMF,i\pm 1}}$ are of order $J_\perp$. Thus,
to lowest order in $J_\perp$ we find that
\begin{equation}
\label{eq:MFcond1}
\overline{\langle n_{\alpha}(\br)\rangle_{\subMF}}=2\rhos J_\perp\int d^2 r'
\overline{\chi(\br,\br')}\;\overline{\langle n_{\alpha}(\br')\rangle_{\subMF}},
\end{equation}
where we used $\overline{\langle n_{\alpha,i}(\br)\rangle_0}=0$ and the independence of
$\overline{\langle n_{\alpha,i}(\br)\rangle_{\subMF,i}}$ and
$\overline{\chi_{\alpha\beta,i}(\br,\br')}=\delta_{\alpha\beta}\overline{\chi(\br,\br')}$
on $i$. For $B=0$ or when disorder or temperature are effective in destroying the Abrikosov lattice,
translational invariance leads to equation (\ref{eq:mf}), which expresses the condition for
the onset of CDW order in term of the in-plane CDW susceptibility
$\overline{\chi(T)}=(1/L^2)\int d^2r d^2r' \overline{\chi(\br,\br')}$.

In a clean system subject to a weak magnetic field at low temperatures, the important contribution
to the integral in equation (\ref{eq:MFcond1}) comes from the vicinity of vortex cores. This implies
an approximate condition for the transition, similar to equation (\ref{eq:mf}), but with the right
hand side multiplied by a factor that scales as the ratio between the core area
$r_0^2\sim 1/g$,\cite{nlsm-vortex} and the area of the magnetic unit cell $R^2\sim \phi_0/B$.

In the presence of interlayer Josephson coupling and in the absence of disorder, the Hamiltonian reads
\begin{equation}
  \label{eq:HJsup}
  H=\sum_i H_0[\psi_i,n_{\alpha,i}]
  -\rhos J_{SC}\int d^2r\sum_{i}\left[\psi^*_{i}\psi_{i+1}+{\rm H.c.}\right],
\end{equation}
where $i$ is the layer index. The interlayer mean-field approximation
amounts to replacing $H$ by an effective single-layer Hamiltonian of
the form
\begin{equation}
  \label{eq:HMFsup}
  H_{\subMF}=H_0-2\rhos J_{SC}\int d^2r\left[\psi^*\langle\psi\rangle_{\subMF}+{\rm H.c.}\right].
\end{equation}

We are interested in using this approximation to estimate $T_c$ in the
multi-layer system.  To this end, we calculate
$\langle\psi\rangle_{\subMF}$. Since it is small in the vicinity of $T_c$
we may carry out the averaging over $H_{\subMF}$ perturbatively in the
Josephson coupling term. As a result, in the absence of a magnetic
field and using the fact that $\langle\psi(\br)\psi(\br')\rangle_0=0$,
we obtain the following condition for the SC transition
\begin{equation}
  \label{eq:condTcsup}
  T_c=2\rhos J_{SC}\int d^2r\langle\psi^*(\br')\psi(\br)\rangle_0.
\end{equation}
For weak $J_{SC}$, $T_c$ lies close to $T_{BKT}$ of a single layer, where
$\langle\psi^*(\br')\psi(\br)\rangle_0\approx|\psi_0(T_c)|^2
\langle e^{i[\theta(\br)-\theta(\br')]}\rangle_0$. Since for $T>T_{BKT}$
the SC phase correlations decay exponentially over the BKT correlation
length $\xi$, we obtain
\begin{equation}
  \label{eq:condTcXYsup}
  T_c=4\pi\rhos J_{SC}|\psi_0(T_c)|^2\xi^2(T_c).
\end{equation}
On a square lattice $T_{BKT}\approx 0.9\rhos|\psi_0(T_{BKT})|^2$, thereby
establishing, for $T_c\approx T_{BKT}$, a relation between $|\psi_0(T_c)|^2$
and $T_{BKT}$. Finally, using BKT critical behavior of
$\xi(T_c)=r_0\exp[b\sqrt{T_{BKT}/(T_c-T_{BKT})}]$,
where $b$ is a constant, we arrive at
\begin{equation}
  \label{MF3dTcsup}
  T_c=T_{BKT}\left[[1+\frac{4b^2}{\ln^2(0.9/4\pi J_{SC}r_0^2)}\right].
\end{equation}

\section{The NLSM with a random pinning potential}
\label{app:disorder}

We next consider the NLSM of a single layer with independent Gaussian
random potentials, $V_\alpha$. The system is described by the action
\begin{eqnarray}
  \label{eq:HVsrcs}
  S & = & \beta H_0 - \beta\rhos\int d^2r\,\sum_\alpha V_\alpha(\br) n_\alpha(\br)
  \nonumber \\  & &  - \int d^2r \sum_\alpha J_\alpha(\br)n_\alpha(\br)
  \nonumber \\ & & - \int d^2rd^2r'\sum_{\alpha\beta}K_{\alpha\beta}(\br,\br')
  n_\alpha(\br)n_\beta(\br'),
\end{eqnarray}
to which we have introduced sources in order to calculate correlation
and response functions in terms of the free energy,
\begin{eqnarray}
  \label{eq:FV}
  e^{-\beta F} & = & \int\D\psi^*\D\psi\D n\,\delta\left(|\psi|^2
    +\sum_{\alpha=1}^{N-2}n_\alpha^2-1\right)e^{-S}.
  \nonumber  \\ & &
\end{eqnarray}
As a result,
\begin{equation}
  \label{eq:G}
  G_{\alpha\beta}(\br,\br')\equiv\overline{\braket{n_\alpha(\br)n_\beta(\br')}}
    =\left.-\frac{\delta\,\beta\overline{F}}{\delta K_{\alpha\beta}(\br,\br')}
      \right|_{K=0}.
\end{equation}
and
\begin{eqnarray}
  \label{eq:chi}
   T\overline{\chi_{\alpha\beta}(\br,\br')} & \equiv & 
    \overline{\braket{n_\alpha(\br)n_\beta(\br')}}
    -\overline{\braket{n_\alpha(\br)}}\overline{\braket{n_\beta(\br')}}
    \nonumber \\ & = & \left.-\frac{\delta^2\,\beta\overline{F}}
    {\delta J_\alpha(\br)\delta J_\beta(\br')}.
  \right|_{J=0}
\end{eqnarray}

The main difficulty is in calculating $\overline{F}$, the free energy
averaged over realizations of disorder. This can be done by employing
the replica method in which we consider $m$ replicas of the original
model. Analytically continuing $m\to 0$ we have
$\overline{F}=\lim_{m\to 0} F(m)/m$, where $F(m)$ is defined by
\begin{widetext}
  \begin{eqnarray}
    \label{eq:Fm}
    \!\!\!\!\!\!e^{-\beta F(m)} & = & \int\D\psi^{a*}\D\psi^{a}\D n_\alpha^a \D V_\alpha
    \delta\left(|\psi^a|^2+\sum_{\alpha=1}^{N-2}(n_\alpha^a)^2-1\right)
    e^{-\sum_a S[\psi^a,n_\alpha^a,J_\alpha,K_{\alpha\beta}]-\frac{1}{2V^2}\int d^2r \sum_\alpha V_\alpha^2}
    \nonumber \\
    \!\!\!\!\!\!\!& = & \int\D\psi^{a*}\D\psi^{a}\D n_\alpha^a \D V_\alpha\D\bar\sigma^a
    e^{-\sum_a S[\psi^a,n_\alpha^a,J_\alpha,K_{\alpha\beta}]-\frac{1}{2V^2}\int d^2r
      \sum_\alpha V_\alpha^2
      +\frac{1}{2}\beta\rhos\sum_a\int d^2r \,i\bar\sigma^a\left[|\psi^a|^2+\sum_\alpha(n_\alpha^a)^2-1\right]}.
  \end{eqnarray}
Integrating over $V_\alpha$ and analytically continuing to
$\bar\sigma^a=i\sigma^a$, we have
  $e^{-\beta F(m)} = \int\D\psi^{a*}\D\psi^{a}\D n_\alpha^a\D\sigma^a e^{-\tilde S(m)}$,
with
\begin{eqnarray}
  \label{eq:Stilde}
  \tilde S(m) &=& \frac{1}{2}\beta\rhos\int d^2r \left\{\sum_a\left[
      |(\bnabla+2ie\bA)\psi^a|^2+\sigma^a(|\psi^a|^2-1)\right]
    +\sum_{ab}\sum_\alpha n_\alpha^a\left[\delta_{ab}\hat L^a
      -\beta\rhos V^2\right]n_\alpha^b\right\} \nonumber\\
      &-&\int d^2r \sum_a\sum_\alpha J_\alpha(\br)n_\alpha^a(\br)
   - \int d^2rd^2r'\sum_{ab}\sum_{\alpha}K_{\alpha\beta}(\br,\br')
  n_\alpha^a(\br)n_\beta^a(\br'),
\end{eqnarray}
and $\hat L^a = -\lambda\nabla^2+g+\sigma^a(\br)$.  Integrating over
the CDW fields, $n_\alpha^a$, gives $e^{-\beta F(m)} =
\int\D\psi^{a*}\D\psi^{a}\D\sigma^ae^{-S(m)}$, where $S(m)$ is defined
by
\begin{eqnarray}
  \label{eq:Sm}
  S(m) & = & \frac{1}{2}\Tr\ln(G^{-1}-2K)+
  \frac{1}{2}\beta\rhos\int d^2r\sum_a\left[
    |(\bnabla+2ie\bA)\psi^a|^2+\sigma^a(|\psi^a|^2-1)\right] \nonumber \\
  & & -\frac{1}{2}\int d^2r\,d^2r'\sum_{ab}\sum_{\alpha\beta}J_{\alpha}(\br)
 \left[\left(G^{-1}-2K\right)^{-1}\right]_{\alpha\beta}^{ab}(\br,\br')J_\beta(\br')
\end{eqnarray}
\end{widetext}
with,
\begin{equation}
  \label{eq:Ginv}
  (G^{-1})_{\alpha\beta}^{ab}(\br,\br') = \beta\rhos[\delta_{ab}\hat L^a-\beta\rhos V^2]
  \delta_{\alpha\beta}\delta(\br-\br'),
\end{equation}
and
\begin{equation}
  \label{eq:K}
  K_{\alpha\beta}^{ab}(\br,\br')=\delta_{ab}K_{\alpha\beta}(\br,\br').
\end{equation}

We would like to calculate the integrals over $\psi^a$, $\psi^{a*}$
and $\sigma^a$, using a saddle-point approximation, which is justified
by $N\to\infty$ for $\sigma^a$, and provided the disorder is weak and satisfies
$V^2a^2={\cal O}(1/N)$, by $\beta\to\infty$ for $\psi^a$ and $\psi^{a*}$,
see Eq. (\ref{eq:psi0V}) below. Within this approximation, $\beta F(m)=S(m)$,
where $S(m)$ is evaluated for the configurations of $\psi^a,\psi^{a*}$
and $\sigma^a$ which solve the saddle-point equations
\begin{equation}
  \label{eq:SPV}
  \frac{\delta\,\beta F(m)}{\delta\sigma^a(\br)}
  = \frac{1}{2}\beta\rhos\left[\sum_\alpha G_{\alpha\alpha}^{aa}(\br,\br)
    +(|\psi^a|^2-1)\right]=0,
\end{equation}
and
\begin{equation}
  \label{eq:SPVpsi}
  \frac{\delta\,\beta F(m)}{\delta\psi^{a*}(\br)} =
  \frac{1}{2}\beta\rhos\left[-(\bnabla+2ie\bA)^2+\sigma^a\right]\psi^a=0.
\end{equation}
Since we are interested in $S(m)$ to order ${\cal O}(J^2,K)$,
these saddle-point equations are not affected by
the source fields $J$ and $K$. Furthermore, since $G^{-1}$ is diagonal
in $\alpha,\beta$ and symmetric in $a,b$ and $\br,\br'$, so is
$G_{\alpha\beta}^{ab} = \delta_{\alpha\beta}G^{ab}$. Thus, we find
\begin{equation}
  \label{eq:Glim}
  G_{\alpha\beta}(\br,\br')=\delta_{\alpha\beta}\lim_{m\to 0}\frac{1}{m}
  \sum_aG^{aa}(\br,\br'),
\end{equation}
and, similarly,
\begin{equation}
  \label{eq:chilim}
  T\overline{\chi_{\alpha\beta}(\br,\br')} = \delta_{\alpha\beta}\lim_{m\to 0}\frac{1}{m}
  \sum_{ab}G^{ab}(\br,\br').
\end{equation}

We will calculate $G^{ab}$ by assuming a replica-symmetric solution of
the saddle-point equations, \ie, $\psi^a=\psi$ and
$\sigma^a=\sigma$. Under this assumption the operator $\hat L^a=\hat L$
is also replica symmetric, and $G^{ab}$ must obey
\begin{equation}
  \label{eq:GGinv}
  \sum_c\beta\rhos(\delta_{ac}\hat L-\beta\rhos V^2)G^{cb}(\br,\br')
  =\delta_{ab}\delta(\br-\br').
\end{equation}
Expanding $G^{aa}$ in the eigen-basis of $\hat L$
\begin{equation}
  \label{eq:Geigen}
  G^{ab}(\br,\br')=\sum_{st}G_{st}^{ab}\phi_s(\br)\phi_t^*(\br'),
\end{equation}
we find the solution
\begin{equation}
  \label{eq:Gsol}
  G_{st}^{ab}=\delta_{st}\left[\frac{\delta_{ab}}{\beta\rhos\varepsilon_s}
    +\frac{V^2}{\varepsilon_s(\varepsilon_s-m\beta\rhos V^2)}\right],
\end{equation}
expressed in terms of the eigenvalues, $\varepsilon_s$, of $\hat L$.

In the absence of a magnetic field, the saddle-point equation for
$\psi$, Eq.  (\ref{eq:SPVpsi}), assumes a uniform solution
$\psi=\psi_0$ with $\sigma=0$.  For this case the spectrum of $\hat L$
is spanned by plane waves $s\equiv\bk$, $\phi_\bk(\br)=e^{i\bk\cdot\br}/L$
and $\varepsilon_\bk=\lambda k^2+g$. Thus, we find for the correlation
function
\begin{eqnarray}
  \label{eq:GV}
  G_{\alpha\beta}(\br,\br') & = & \delta_{\alpha\beta}\lim_{m\to 0}
  \frac{1}{m}\sum_a\sum_\bk G_\bk^{aa}\frac{1}{L^2}e^{i\bk\cdot(\br-\br')}
    \nonumber \\ & = & \delta_{\alpha\beta}\int\frac{d^2k}{(2\pi)^2}
    \left[\frac{1}{\beta\rhos(\lambda k^2+g)}+\frac{V^2}
      {(\lambda k^2+g)^2}\right] \nonumber \\ & &
    \qquad\qquad\qquad\qquad\qquad \times e^{i\bk\cdot(\br-\br')},
\end{eqnarray}
from which follows the averaged structure factor
\begin{eqnarray}
  \label{eq:SV}
  \nonumber
  \overline{S(\bq)} &=& \frac{1}{L^2}\int d^2r\,d^2r'\,
  e^{-i\bq\cdot(\br-\br')}G_{\alpha\alpha}(\br,\br') \\
  & = & \frac{T}{\rhos (\lambda q^2+g)}+\frac{V^2}{(\lambda q^2+g)^2}.
\end{eqnarray}
Similarly, we find that the response function is given by
\begin{eqnarray}
  \label{eq:chiV}
  T\overline{\chi_{\alpha\beta}(\br,\br')} & = & \delta_{\alpha\beta}\lim_{m\to 0}
  \frac{1}{m}\sum_{ab}\sum_{\bk}G_\bk^{ab}\frac{1}{L^2}e^{i\bk\cdot(\br-\br')}
  \nonumber \\ & = &   \delta_{\alpha\beta}\frac{1}{\beta\rhos}
  \int\frac{d^2k}{(2\pi)^2}\frac{e^{i\bk\cdot(\br-\br')}}{\lambda k^2+g},
\end{eqnarray}
such that the susceptibility is
\begin{eqnarray}
  \label{eq:chi0}
  \nonumber
  \overline{\chi(\bq)}&=&\frac{1}{L^2}\int d^2r\,d^2r'\, e^{-i\bq\cdot(\br-\br')}
  \overline{\chi_{\alpha\alpha}(\br,\br')} \\
  &=&\frac{1}{\rhos(\lambda q^2+ g)}.
\end{eqnarray}
Note that unlike the clean case, disorder-induced correlations between neighboring
regions lead to $\overline{\langle n_\alpha(\br)\rangle\langle n_\alpha(\br')\rangle}\neq 0$,
and therefore to $\overline{\chi(\bq)}\neq \overline{S(\bq)}/T$.

Using Eqs. (\ref{eq:Geigen}) and (\ref{eq:Gsol}) we obtain that in the $m\to 0$ limit
the saddle-point equation for $\sigma$, Eq. (\ref{eq:SPV}), takes the form
\begin{equation}
  \int\frac{d^2k}{(2\pi)^2}
  \left[\frac{1}{\beta\rhos(\lambda k^2+g)}+\frac{V^2}
    {(\lambda k^2+g)^2}\right]  = \frac{1-|\psi_0|^2}{N-2},
\end{equation}
which, for $\lambda\gg ga^2$, gives
\begin{equation}
  \label{eq:psi0V}
  |\psi_0|^2\approx 1-\frac{(N-2)V^2}{4\pi\lambda g}-\frac{T}{T_{MF}^0},
\end{equation}
where $T_{MF}^0$ is the value of $T_{MF}$ in the clean system. We
therefore find that disorder reduces $|\psi_0|^2$, as well as
$T_{MF}$. Note that the solution, Eq. (\ref{eq:psi0V}), exists only for
weak enough disorder. For stronger disorder the saddle-point configuration
is $\psi=0$ and $\sigma>0$, thus indicating the need to take into account
fluctuations in $\psi$.

Next, let us include the effects of a magnetic field on $\overline{S_{CDW}}$
and $\overline{\chi}$. Just as for the clean
system, we expect that the saddle-point equations of the replicated
action possess a solution in the form of an Abrikosov lattice. Hence,
we assume that the spectrum of $\hat L$ consists of a continuum of
scattering states, similar to those of the magnetic-field-free system,
and a band originating from bound states inside vortex cores. The
reasoning that was used for the derivation of Eq. (\ref{eq:spbound})
is then applicable here.  If, in addition, we assume that
$t\ll\varepsilon_0$, we can ignore the dispersion of the tight-binding
band and approximate it by a flat band with eigenvalue $\varepsilon_0$.
Consequently, the saddle-point equation for $\sigma$ becomes, as $m\to 0$
\begin{eqnarray}
  \label{eq:SPVAL}
  &&\hspace{-1cm}\sum_\bR\left[\frac{1}{\beta\rhos\varepsilon_0}+\frac{V^2}
    {\varepsilon_0^2}\right]|\varphi_0(\br-\bR)|^2 \nonumber \\
    &&\hspace{2cm} \approx \frac{
    \left[|\psi_0|^2-|\psi(\br)|^2+|\delta\psi(\br)|^2\right]}{N-2}.
\end{eqnarray}
Integrating over $\br$ and dividing by the system area gives
\begin{equation}
  \label{eq:SPVALavg}
  \frac{1}{R^2}\left[\frac{1}{\beta\rhos\varepsilon_0}+\frac{V^2}{\varepsilon_0^2}
    \right] = {\cal C}|\psi_0|^2\left(\frac{r_0}{R}\right)^2,
\end{equation}
where ${\cal C}$ is a numerical constant.
From Eq. (\ref{eq:SPVALavg}) we find that the assumption $t\ll \varepsilon_0$
is indeed satisfied under reasonable conditions, \ie, $t\ll \max (gT/\rhos,\sqrt{gV^2})$.

In the presence of disorder the expansion of the
structure factor in terms of the eigenstates and eigenvalues of
$\hat L$ takes the form
\begin{equation}
  \label{eq:SVB}
  \overline{S(\bq)} = \frac{1}{L^2}\sum_s\left[\frac{1}{\beta\rhos\varepsilon_s}
    +\frac{V^2}{\varepsilon_s^2}\right]\left|\int d^2r\,
    e^{-i\bq\cdot\br}\phi_s(\br)\right|^2.
\end{equation}
Due to the same consideration used for the clean systems, we find that
the main contribution of the scattering states comes from the lowest
lying state with $\varepsilon_s\simeq g$. An additional contribution
comes from the states $\varphi_0(\br-\bR)$ bound to the vortex cores at
positions $\bR$. Noting that $L^{-2}\sum_\bR|\int d^2r\varphi_0(\br-\bR)|^2\sim
1/(gR^2)$, we obtain Eq. (\ref{eq:S5}).

For the response function one finds
\begin{equation}
  \label{eq:chiVB}
  \overline{\chi(\bq)} = \frac{1}{L^2}\sum_s\frac{1}{\rhos\varepsilon_s}
    \left|\int d^2r\, e^{-i\bq\cdot\br}\phi_s(\br)\right|^2,
\end{equation}
yielding for $T\to 0$ and $t\ll\sqrt{gV^2}$ a $\bq=0$ susceptibility
\begin{equation}
  \label{eq:chisusVB}
  \overline{\chi} = \frac{1}{\rhos g}\left(1+{\widetilde{\cal C}}\frac{B}{\phi_0\sqrt{gV^2}}\right),
\end{equation}
with $\widetilde{{\cal C}}$ a constant. Finally, the $T\to 0$ spatially averaged
Edwards-Anderson order parameter can be easily calculated from the saddle-point equation,
Eq. (\ref{eq:SPVAL}),
\begin{eqnarray}
  \label{eq:qEA0}
  q_{EA}(T\to 0) & = & \frac{1}{L^2}\int d^2r\, G_{\alpha\alpha}(\br,\br)
  \nonumber \\ & = & \frac{1 - |\psi_0|^2+{\cal C}|\psi_0|^2
    \left(\frac{r_0}{R}\right)^2}{N-2},
\end{eqnarray}
which gives Eq. (\ref{eq:qEA}).

\end{document}